\documentclass[twocolumn,showpacs,superscriptaddress,amsmath,amssymb]{revtex4}

\usepackage{epsfig}
\usepackage{graphicx}
\usepackage{dcolumn}
\usepackage{bm}

\begin{document}


\title{Pion Production by Protons on a Thin Beryllium Target \\ at 6.4, 12.3, and 17.5 GeV/c Incident Proton Momenta}

\date{\today}

\affiliation{Brookhaven National Laboratory, Upton, NY 11973}
\affiliation{Columbia University, New York, NY 10027}
\affiliation{Florida State University, Tallahassee, FL 32306}
\affiliation{Illinois Institute of Technology, Chicago, IL 60616}
\affiliation{Iowa State University, Ames, IA 50010}
\affiliation{Kent State University, Kent, OH 44242}
\affiliation{Laboratori Nazionali di Frascati dell'INFN, Frascati, Italy I-00044}
\affiliation{Lawrence Berkeley National Laboratory, Berkeley, CA 94720}
\affiliation{Lawrence Livermore National Laboratory, Livermore, CA 94550}
\affiliation{Oak Ridge National Laboratory, Oak Ridge, TN 37831}
\affiliation{State University of New York, Stony Brook, NY 11794}
\affiliation{Texas A\&M University,  College Station, TX 77843}
\affiliation{University of Illinois, Urbana-Champaign, Illinois 61801}
\affiliation{University of Tennessee, Knoxville, TN 37996}
\affiliation{Virginia Polytechnic Institute and State University, Blacksburg, VA 24061}
\affiliation{Yonsei University, Seoul 120-749, Korea}

\author{I.~Chemakin}
\affiliation{Columbia University, New York, NY 10027}
\author{V.~Cianciolo}
\affiliation{Lawrence Livermore National Laboratory, Livermore, CA 94550}
\affiliation{Oak Ridge National Laboratory, Oak Ridge, TN 37831}
\author{B.~A.~Cole}
\affiliation{Columbia University, New York, NY 10027}
\author{R.~C.~Fernow}
\affiliation{Brookhaven National Laboratory, Upton, NY 11973}
\author{A.~D.~Frawley}
\affiliation{Florida State University, Tallahassee, FL 32306}
\author{M.~Gilkes}
\affiliation{State University of New York, Stony Brook, NY 11794}
\author{S.~Gushue}
\affiliation{Brookhaven National Laboratory, Upton, NY 11973}
\author{E.~P.~Hartouni}
\affiliation{Lawrence Livermore National Laboratory, Livermore, CA 94550}
\author{H.~Hiejima}
\affiliation{Columbia University, New York, NY 10027}
\affiliation{University of Illinois, Urbana-Champaign, Illinois 61801}
\author{M.~Justice}
\affiliation{Kent State University, Kent, OH 44242}
\author{J.~H.~Kang}
\affiliation{Yonsei University, Seoul 120-749, Korea}
\author{H.~G.~Kirk}
\affiliation{Brookhaven National Laboratory, Upton, NY 11973}
\author{J. M. Link}
\email{Jonathan.Link@vt.edu}
\affiliation{Columbia University, New York, NY 10027}
\affiliation{Virginia Polytechnic Institute and State University, Blacksburg, VA 24061}
\author{N.~Maeda}
\affiliation{Florida State University, Tallahassee, FL 32306}
\author{R.~L.~McGrath}
\affiliation{State University of New York, Stony Brook, NY 11794}
\author{S.~Mioduszewski}
\affiliation{University of Tennessee, Knoxville, TN 37996}
\affiliation{Brookhaven National Laboratory, Upton, NY 11973}
\affiliation{Texas A\&M University,  College Station, TX 77843}
\author{J. Monroe}
\affiliation{Columbia University, New York, NY 10027}
\author{D.~Morrison}
\affiliation{University of Tennessee, Knoxville, TN 37996}
\affiliation{Brookhaven National Laboratory, Upton, NY 11973}
\author{M.~Moulson}
\affiliation{Columbia University, New York, NY 10027}
\affiliation{Laboratori Nazionali di Frascati dell'INFN, Frascati, Italy I-00044}
\author{M.~N.~Namboodiri}
\affiliation{Lawrence Livermore National Laboratory, Livermore, CA 94550}
\author{G.~Rai}
\affiliation{Lawrence Berkeley National Laboratory, Berkeley, CA 94720}
\author{K.~Read}
\affiliation{University of Tennessee, Knoxville, TN 37996}
\author{L.~Remsberg}
\affiliation{Brookhaven National Laboratory, Upton, NY 11973}
\author{M.~Rosati}
\affiliation{Brookhaven National Laboratory, Upton, NY 11973}
\affiliation{Iowa State University, Ames, IA 50010}
\author{Y.~Shin}
\affiliation{Yonsei University, Seoul 120-749, Korea}
\author{R.~A.~Soltz}
\affiliation{Lawrence Livermore National Laboratory, Livermore, CA 94550}
\author{M. Sorel}
\affiliation{Columbia University, New York, NY 10027}
\author{S.~Sorensen}
\affiliation{University of Tennessee, Knoxville, TN 37996}
\author{J.~H.~Thomas}
\affiliation{Lawrence Livermore National Laboratory, Livermore, CA 94550}
\affiliation{Lawrence Berkeley National Laboratory, Berkeley, CA 94720}
\affiliation{Brookhaven National Laboratory, Upton, NY 11973}
\author{Y.~Torun}
\affiliation{State University of New York, Stony Brook, NY 11794}
\affiliation{Brookhaven National Laboratory, Upton, NY 11973}
\affiliation{Illinois Institute of Technology, Chicago, IL 60616}
\author{D.~L.~Winter}
\affiliation{Columbia University, New York, NY 10027}
\author{X.~Yang}
\affiliation{Columbia University, New York, NY 10027}
\author{W.~A.~Zajc}
\affiliation{Columbia University, New York, NY 10027}
\author{Y.~Zhang}
\affiliation{Columbia University, New York, NY 10027}

\begin{abstract}
An analysis of inclusive pion production in proton-beryllium collisions at 6.4, 12.3, and 17.5 GeV/c proton beam momentum has been performed.  The data were taken by Experiment 910 at the Alternating Gradient Synchrotron at the Brookhaven National Laboratory.  The differential $\pi^+$ and $\pi^-$ production cross sections ($d^2\sigma/dpd\Omega$) are measured up to 400~mrad in $\theta_{\pi}$ and up to 6~GeV/c in $p_{\pi}$.  The measured cross section is fit with a Sanford-Wang parameterization.
\end{abstract}

\maketitle

\section{\label{introduction}Introduction}
A detailed understanding of the production of pions in proton interactions with nuclear targets is essential for determining the flux of neutrinos in accelerator based neutrino experiments.  Flux predictions are particularly difficult for experiments using lower energy primary proton beams such as MiniBooNE~\cite{MiniBooNE} (8~GeV) and K2K~\cite{Ahn:2001cq} (12~GeV), where there currently exist large uncertainties in the pion production cross section data.  Most of the existing data with proton beam energies in the 5 to 20~GeV range were taken over 30 years ago using single arm spectrometers~\cite{Dekkers:1965, Asbury:1969bf, Cho:1972jq, Baker:1961, Marmer:1969if, Lundy:1965}, but more recently the HARP Experiment at CERN has started to publish new data~\cite{harp:2007gt}.  A global fit to the older data by Cho {\em et al.}~\cite{Cho:1972jq} found a normalization discrepancy of $\sim$15\% between the various experiments and more recent fits~\cite{Monroe:2004xe,Hill:2001pu} have also found general inconsistencies. 

The high statistics data taken by Brookhaven experiment 910 (E910) provides an opportunity to revisit these old measurements with a modern, wide angle spectrometer.  The experiment covers a wide range of secondary momenta and angles with particle identification over most of this range. 

E910 has previously published $\pi^{\pm}$ production cross sections for low momentum pions (0.1 to 1.2~GeV/c) on several different target materials including beryllium~\cite{Chemakin:2001qx}.  This analysis extends that earlier work to higher pion momenta.

\section{\label{experiment}The Experiment}
E910 ran for 14 weeks at the MPS facility in the A1 secondary beam line of the BNL AGS in 1996.  The total momenta and directions of incoming beam protons were reconstructed using proportional chambers located upstream of the target.  Data were taken at three beam momenta: 6.4, 12.3, and 17.5~GeV/c.  Beam protons were identified by three \v{C}erenkov counters along the beam line.

\begin{figure}
\includegraphics[width=8.cm]{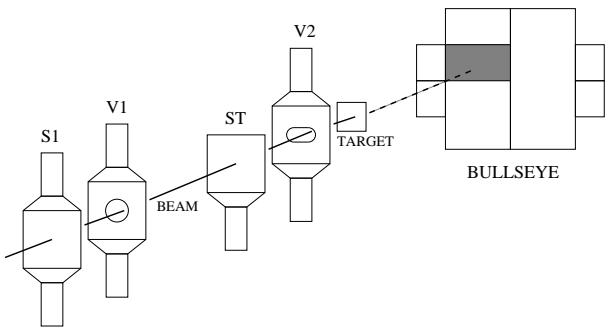}
\caption{\label{trigger}The beam trigger counters and bullseye counter.  The shaded area on the bullseye defines the beam veto region.  A hit in this region indicates that the event is consistent with a non-interacting beam particle.}
\end{figure}

A set of trigger counters ($S1$, $ST$) and veto counters ($V1$, $V2$) located between the proportional chambers and the target were used to detect and constrain the trajectories of incoming beam particles.  The trigger counters are shown in Figure~\ref{trigger}.  The trigger configuration for the data used in this analysis employs the ``bullseye'' counter, which was located 6.8~m downstream of the target.  The bullseye counter consists of two pairs of scintillator slats, one pair of 14.6$\times$30.5~cm slats placed along the vertical and the second of 40.6$\times$7.6~cm slats aligned horizontally.  Non-interacted beam particles consistent with the aperture defined by the veto counters are entirely located within the intersection of one of the horizontal and one of the vertical slats as shown in Figure~\ref{trigger}.  The minimum bias trigger for this analysis, known as the bullseye trigger, is defined to be the combination of the presence of beam ($S1\cdot\overline{V1}\cdot ST\cdot\overline{V2}$) and the absence of a hit in the relevant slats of the bullseye counter.

During the run, a variety of target materials were used.  This analysis focuses on the proton-beryllium (Be) interaction data sets.  In the rest of the paper, only Be target data will be discussed.  The Be target had a geometric cross section of 7.62$\times$2.54~cm$^2$ and was 1.84$\pm$0.04~cm long ($\sim$ 4.5\% of an interaction length).  The beam spot on the target was defined by the last veto scintillator, which had a 2$\times$1~cm$^2$ slot with semicircular ends.

\begin{figure}
\includegraphics[width=9.cm]{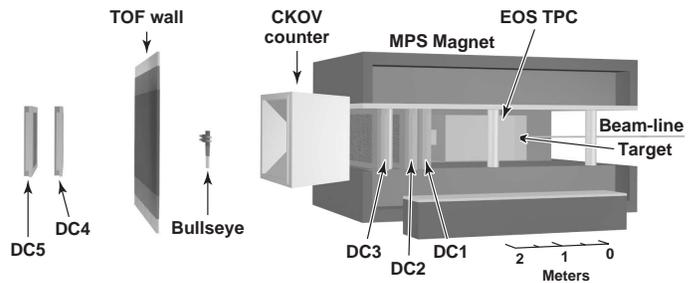}
\caption{\label{spect}The E910 spectrometer layout.}
\end{figure}

Reaction products from proton-beryllium interactions were measured with the spectrometer layout shown in Figure~\ref{spect}.  The target was located 10~cm upstream of the time projection chamber (TPC) active volume.  The EOS TPC~\cite{Rai:1990jp} is 1.54~m long and is read out through a 120$\times$128 cathode pad array.  It ran with P10 gas at atmospheric pressure and a vertical electric field of 120~V/cm.  The TPC was located in the center of the MPS magnet, which had a nominal central field of 0.5~T along the vertical axis.  Downstream of the TPC, inside the magnet, charged particle tracking was provided by three drift chambers (DC1-3).  Each drift chamber consisted of seven wire planes: three $x$ views (one staggered), two $y$ views (staggered), and two views rotated from vertical by $\pm60^{\circ}$.  A segmented threshold \v{C}erenkov counter, with an aperture of 139.7$\times$190.5~cm$^2$, was located 4.8~m downstream of the target.  The counter used 96 separate mirrors, a central 8$\times$8 grid of small mirrors surrounded by 32 mirrors with a factor of four larger aperture, to direct \v{C}erenkov light produced by particles traversing the nominally 1~m length of the counter to the same number of photomultiplier tubes located at the top of the counter.  The radiating medium was Freon 114.  The time-of-flight (TOF) wall was located 8~m downstream of the target.  It consisted of 32 scintillating counter slats, each 15.2$\times$178~cm$^2$, arrayed in a flat panel of approximately 488$\times$178~cm$^2$.  The typical TOF resolution was  $\sim$160~ps~\cite{Chemakin:2001vw}.  Two more drift chambers (DC4-5) were located beyond the TOF wall at 9.6 and 10.1~m from the target.

\section{Data Analysis}
The differential cross section for pion production as a function of pion momentum, $p$, and zenith angle, $\theta$, is given in the lab frame by 
\begin{equation*}
\frac{d^2\sigma}{dpd\Omega}(p,\theta) = \frac{A}{N_A \ell \rho}\ \frac{1}{\varepsilon}\ \frac{1}{a(p,\theta)}\ \frac{1}{\Delta p 2\pi\Delta \cos \theta}\  \frac{N_{\pi}(p,\theta)}{N_{beam}}
\end{equation*}
where $A$ is the target mass number ($A_{Be}=9.01~g/mole$), $N_A$ is Avogadro's number, $\ell$ is the target length, $\rho$ is the target density ($\rho_{Be}=1.848$~g/cm$^3$), $\varepsilon$ is the trigger efficiency, $a(p,\theta)$ is the geometrical acceptance and cut efficiency measured (with Monte Carlo simulation) as a function of $p$ and $\theta$, $\Delta p 2\pi \Delta\cos\theta$ is the  area of the bin in ($p\,\Omega$) space, $N_{\pi}(p,\theta)$ is the number of pions observed in the bin, and $N_{beam}$ is the total number of protons incident on the target.  

The data are binned in 400~MeV/c wide bins in momentum.  In the 12.3 and 17.5~GeV/c data sets, the $\theta$ bin width is 60~mrad from 0 to 360~mrad, and for the 6.4~GeV/c data set, the $\theta$ binning is 100~mrad from 0 to 400~mrad to accommodate lower statistics.
  
\subsection{Event Selection}
For this analysis, all events - both those used to determine $N_{beam}$ and $N_{\pi}$ - must satisfy the following criteria: the beam particle must be successfully tracked in the beam proportional chambers; the beam \v{C}erenkov response must be consistent with the proton hypothesis; the reconstructed primary vertex must be consistent with the target $z$ position and lie within the open apertur of the $V2$ counter.  

A completely unbiased beam trigger is used to determine the the number of beam protons, $N_{beam}$.  This trigger fired on the presence of beam only, without requiring an interaction.  The trigger was heavily prescaled; by factors of 32 to 64 depending on the run.  The number of beam protons is determined by counting all beam trigger events and multiplying by the prescale factor.  As a cross check on the prescale, the prescale factor is computed in the data by taking the ratio of bullseye trigger events which also have a prescaled beam trigger to all bullseye triggers.  In all data sets this number agrees with the input prescale to well within the statistical error on the ratio.

Candidate events are required to have a bullseye trigger, and a reconstructed vertex position consistant with the reconstructed beam track and within the volume of the target.  The reconstructed secondaries must be consistent with a single inelastic collision (i.e. not consistent with an elastically scattered beam particle and the sum of the secondary particle momenta must not be greater than the beam momentum).

\subsection{Track Selection}
All candidate tracks must point back to the primary interaction vertex, they must have a minimum of 20 hits in the TPC, and the TPC $dE/dx$ calculation must have been successful.

\subsection{Trigger Efficiency}
The bullseye trigger efficiency, $\varepsilon$, was determined by using a sample of totally unbiased beam trigger events.  The denominator of the efficiency is the number of beam trigger events with at least one secondary track, and the numerator is the subset of those events that also have a bullseye trigger.  The measured trigger efficiency for each beam momentum data set is listed in Table~\ref{stats}.  The error on the trigger efficiency is due only to statistics.

In addition to the flat trigger inefficiency, which is largely due to other tracks in the event that pass through the bullseye, we studied the possible inefficiency as a function of pion kinematics ({\em i.e.} for a particular angle and momentum the track in question may have a non-zero probability of passing through the bullseye).  This effect was determined to be small ($\leq$2\%) in all bins and mostly affects the lowest angle bin.  The bin-to-bin inefficiency is accounted for as a systematic on a bin-by-bin basis.

\begin{table}[tb]
\centerline{
\begin{tabular}{|l|c|c|} \hline
\multicolumn{1}{|c|}{Data Set} & \multicolumn{1}{|c|}{Beam Protons} & \multicolumn{1}{|c|}{Trigger Efficiency}  \\ \hline
6.4 GeV/c  & 93,632    & 100.$\pm$1.1\%   \\ \hline
12.3 GeV/c & 745,216   & 96.8$\pm$0.6\%   \\ \hline
17.5 GeV/c & 2,576,352 & 89.6$\pm$0.6\%   \\ \hline
\end{tabular}}
\caption{\label{stats}Number of protons on target and trigger efficiency for each data set.} 
\end{table}

\subsection{Particle Identification}
Three particle identification (PID) systems are used to distinguish pions from other secondaries in this analysis: the TPC, time-of-flight wall, and threshold \v{C}erenkov counter.  The TPC uses energy loss ($dE/dx$) to distinguish between different particles types.  The pion threshold in \v{C}erenkov counter is about 2.5~GeV/c while the kaon and proton thresholds are both outside the momentum range of this analysis (9.0 and 17.1 respectively).  The TOF wall performance is discussed in ref~\cite{Chemakin:2001vw}. 

For each of the PID detectors, a residual is formed between the true response and the expected response for the different particle hypotheses ($e$, $\pi$, $K$, $p$).  By construction, the residual for the correct hypothesis is Gaussian distributed with a mean of zero and a width of one.  

Pion candidate tracks are divided into three groups as a function of momentum.  The groups are defined by the capabilities of the three PID systems.  In each momentum region the pion residual for the primary PID system is plotted for all candidate track in each ($p,\,\theta$) bin.  These residual distributions are then fit to a unit Gaussian to determine the pion yields.  Information from the other two PID systems is used as a discrete cut (described below) where applicable.

In the range of 0.4 to 1.2~GeV/c the TPC $dE/dx$ is used as the primary PID system (the $dE/dx$ distributions for different particle types can be found in Figure~3 of Ref.~\cite{Chemakin:2001qx}).  The pion-proton and pion-electron separation is excellent over this entire range.  The pion-kaon residuals start to overlap at the 3~$\sigma$ level by about 0.8~GeV/c.  Kaon production is smaller than pion production by a factor of $\sim$20, and the residual overlap is not complete.  Therefore, kaon contamination represents an error on the pion yield of much less than 5\%.  The assertion that the kaon contamination is small is affirmed by the quality of the fits to the pion residuals.  The TOF PID system has a small geometrical acceptance in this range, but if TOF information is available for a track, the TOF pion residual is required to be within $\pm3\:\sigma$ of zero.  The \v{C}erenkov system has no ability to separate pions from kaons and protons in this region and therefore is not used.

Above 1.2~GeV/c, the primary PID system is the TOF.  In the TOF system pion-kaon separation is good ($\geq\,3\,\sigma$) up to $\sim$3~GeV/c and pion-proton separation is good up to $\sim$5.4~GeV/c.  Above 2.8~GeV/c all pions should produce a robust \v{C}erenkov signal, while kaons and protons do not.  In this region, the \v{C}erenkov pion residual must be within 3~$\sigma$ of zero for all pion candidate tracks.  Both below and above 2.8~GeV/c, there is a requirement that TPC $dE/dx$ pion residual be within 3~$\sigma$ of zero.  Below 2.8~GeV/c, this cut is useful for separating pions from electrons.  Above $\sim$3~GeV/c, the  relativistic rise of the pion in $dE/dx$ provides additional pion-proton separation.

\subsection{Acceptance and Efficiency}
A large sample of Monte Carlo events ($\sim$680,000) was used to determine the pion geometric acceptance and cut efficiency, by taking the ratio of generated to accepted $\pi^+$ and $\pi^-$ in each ($p,\,\theta$) bin.  The product of acceptance and reconstruction efficiency was determined by binning in the {\it generated} particle momenta.  The effects of finite momentum and angular resolution on the reconstructed spectra were separately evaluated as systematic errors.  On a bin-by-bin basis, the size of the smearing uncertainty is generally less than 5\%.
 
\section{Results}
The inclusive pion production cross section in proton-beryllium interactions is calculated from candidate pion tracks, binned in $p$ and $\theta$.  Each momentum bin spans 0.4~GeV/c.  The first bin begins at 0.4~GeV/c and the final bin ends at 5.6~GeV/c.  Six zenith angle bins, ranging from 0 to 360~mrad, are used with the 12.3 and 17.5~GeV data sets.  For the much smaller 6.4 GeV data set only four angular bins are used covering the range 0 to 400~mrad.  The angular bin spacing is uniform in $\theta$, and the bin centers are reported as the average $\cos\theta$ of the bin.  Tables~\ref{data_6}, \ref{data_12} and \ref{data_17} list the measured value of the $\pi^+$ and $\pi^-$ cross sections in each ($p,\,\theta$) bin, in units of mbarns/[(GeV/c) steradian].  

The errors reported in Tables~\ref{data_6}, \ref{data_12} and \ref{data_17} include contributions from data statistics, Monte Carlo statistics and bin-by-bin cross checks. These checks include: PID studies, bin migration studies, and a study of bullseye trigger inefficiency as a function of $p$ and $\theta$.  On average, the largest systematic contribution comes from a comparison of the cross section of $\pi^-$ production measured with and without the PID cuts.  To first approximation, all negative secondary tracks are pions.  Therefore, it is possible to measure the $\pi^-$ cross section without using PID cuts.  This PID-free cross section should agree well with the full PID based analysis, and any areas where the PID cut efficiency, as calculated in the Monte Carlo, is not a perfect match to the data would be highlighted.  One expects that deviations might appear in the transition regions such as just above 1.2~GeV/c where the primary PID system switches between TPC and TOF, and around the pion threshold in the \v{C}erenkov counter at $\sim$2.8~GeV/c.  Bin-for-bin the difference between the two analyses is taken as a measure of PID systematic error which is applied to both the $\pi^-$ and $\pi^+$ error analysis.  The PID systematic is typically less 5\%, but can be larger in the PID overlap regions.

The error in each bin (reported in Tables~\ref{data_6}, \ref{data_12} and \ref{data_17}) is the quadratic sum of all systematic contributions and the statistical error.  In addition, there is an overall normalization uncertainty which should be applied equally to all bins.  This error is due in part to a 2\% uncertainty on the measurement of the target thickness, and to the uncertainty in the trigger efficiency shown in Table~\ref{stats}.  The total normalization error is estimated to be less than 5\%.

\section{Fit to the Data}
Modeling the pion production cross section as a function of beam momentum, secondary particle momentum, and secondary particle angle is of interest for input to Monte Carlo simulations.  For this purpose, we fit the results of the previous section with a Sanford-Wang function\cite{Wang:1970bn}, which describes the inclusive double differential pion production cross section in proton-beryllium interactions.  

Sanford-Wang fits have been used in recent years to describe the inclusive pion production cross section for low energy neutrino experiments \cite{Monroe:2004xe}, \cite{Hill:2001pu}, and in the more distant past for global fits to low energy inclusive pion production data \cite{Cho:1972jq}.  The functional form of the parameterization was developed empirically, based on data with incident proton momenta between 10 and 70 GeV/c, therefore fitting the 6.4 GeV/c data provides a useful test of its range of validity.  The explicit form of the Sanford-Wang parameterization used in this analysis is
\begin{eqnarray}
\label{swequation}
SW & = & \mathbf{c_1} p_{\pi}^{\mathbf{c_2}} \left(1 - \frac{p_{\pi}}{p_b-1}\right) \times \\
& & \exp\!\left(\!\frac{-\mathbf{c_3}\,p_{\pi}^{\mathbf{c_4}}}
{p_b^{\mathbf{c_5}}}\!-\!\mathbf{c_6}\, \theta_{\pi}(p_{\pi}\!-\! \mathbf{c_7}\,p_b \cos^{\mathbf{c_8}}\!\theta_{\pi})\!\right) \nonumber
\end{eqnarray}
where $p_{\pi}$ is the momentum of the pion (in GeV/c), $p_b$ is the momentum of the beam proton (in GeV/c), $\theta_{\pi}$ is the production angle (in radians) of the pion in the lab frame, and the $\mathbf{c_i}$ are parameters to be obtained by a fit to the data.  

To fit for the parameters, $\mathbf{c_1}$ through $\mathbf{c_8}$, we use the following $\chi^2$ function:
\begin{equation}
\chi^2 \ = \ \sum_{i,j} \frac{\left(N_j \times \overline{SW}_i - \left(\frac{d^2\sigma}{dpd\Omega}\right)_i\right)^2}{\sigma_i^2}  + \sum_j\frac{(1-N_j)^2}{\sigma_{N_j}^2}
\end{equation}
where $i$ spans all data points in $p_{\pi}$ and $\theta_{\pi}$, and $j$ spans the three beam momentum settings; $N_j$ is a normalization term for the $j$th beam momentum; $\overline{SW}_i$ is the Sanford-Wang function (Eqn~\ref{swequation}) averaged over $p_{\pi}$ and $\theta_{\pi}$ in the $i$th bin:
\begin{eqnarray}
\overline{SW}_i & = &  
\frac{1}{\Delta p_{\pi} 2\pi\Delta \cos\theta_{\pi}} \times \\
 & & \int_{p_i^{\rm lo}}^{p_i^{\rm hi}}\!
\int_{\theta_i^{\rm lo}}^{\theta_i^{\rm hi}} SW(p_b,p_{\pi},\theta_{\pi})
               \sin\theta_{\pi} dp_{\pi}d\theta_{\pi} \nonumber;
\end{eqnarray}
$(d^2\sigma/dpd\Omega)_i$ is the measured cross section in the $i$th bin, and $\sigma_i$ is the measurement error on bin $i$ including both systematic and statistical errors (the normalization error is not included).  The normalization uncertainty of each incident proton momentum data set is handled with the terms $N$, which add to the $\chi^2$ relative to the normalization error, $\sigma_N$.   For all incident proton momentum data sets, $\sigma_N$ is taken to be 5\%.  

\begin{table}[b]
\centerline{
\begin{tabular}{|l|c|c|} \hline
\multicolumn{1}{|c|}{Parameter} & \multicolumn{1}{|c|}{$\pi^+$ Data Fit} & \multicolumn{1}{|c|}{$\pi^-$ Data Fit} \\
\hline
c1       &      258.2      &        249.3      \\ \hline
c2       &      1.018      &        1.066      \\ \hline
c3       &      2.953      &        3.311      \\ \hline
c4       &      2.204      &        1.188      \\ \hline
c5       &      1.782      &        1.017      \\ \hline
c6       &      5.136      &        5.127      \\ \hline
c7       & $7.706 \times 10^{-2}$  &  $6.459 \times 10^{-2}$  \\ \hline
c8       &      14.64      &        10.22      \\ \hline
\hline
$\chi^2$     & 323 & 268    \\ \hline
$N_{d.o.f.}$ & 167 & 169    \\ \hline
\end{tabular}}
\caption{\label{swfit_1}Best-fit parameters for Sanford-Wang fits to the combined 6.4, 12.3, and 17.5~GeV/c production data sets in $\pi^+$ and $\pi^-$.  }
\end{table}

The best fit parameters for the Sanford-Wang fit to the combined 6.4, 12.3, and 17.5~GeV/c $\pi^+$ ($\pi^-$) data are shown in Table~\ref{swfit_1}.  The $\chi^2$ per degree of freedom for the $\pi^+$ ($\pi^-$) fit is 1.93 (1.59).  For both fits, the values of the normalization terms at the best fit point are within 1~$\sigma$ of zero: 1.03 (1.01), 1.01 (1.01), and 0.97 (0.98), for the $\pi^+$ ($\pi^-$) fit in order of increasing beam momentum.  The parameters are highly correlated, so their errors are given as a full covariance matrix, which for the $\pi^+$ fit is: 

\begin{widetext}
\begin{equation*}
\left(\!\begin{array}{cccccccc}
210. & 0.587 & 10.6 & 5.69 & 1.37 & 6.36 & 0.124 & 57.0 \\
0.587 & 1.67\!\times\! 10^{-3} & 2.98\!\times\! 10^{-2} & 1.60\!\times\! 10^{-2} & 3.85\!\times\! 10^{-3} & 1.79\!\times\! 10^{-2} & 3.48\!\times\! 10^{-4} & 0.160 \\
10.6 & 2.98\!\times\! 10^{-2} & 0.541 & 0.289 & 6.97\!\times\! 10^{-2} & 0.323 & 6.30\!\times\! 10^{-3} & 2.89 \\
5.69 & 1.60\!\times\! 10^{-2} & 0.289 & 0.156 & 3.74\!\times\! 10^{-2} & 0.173 & 3.38\!\times\! 10^{-3} & 1.55 \\
1.37 & 3.85\!\times\! 10^{-3} & 6.97\!\times\! 10^{-2} & 3.74\!\times\! 10^{-2} & 9.04\!\times\! 10^{-3} & 4.17\!\times\! 10^{-2} & 8.14\!\times\! 10^{-4} & 0.374 \\
6.36 & 1.79\!\times\! 10^{-2} & 0.323 & 0.173 & 4.17\!\times\! 10^{-2} & 0.195 & 3.77\!\times\! 10^{-3} & 1.73 \\
0.124 & 3.48\!\times\! 10^{-4} & 6.30\!\times\! 10^{-3} & 3.38\!\times\! 10^{-3} & 8.14\!\times\! 10^{-4} & 3.77\!\times\! 10^{-3} & 7.40\!\times\! 10^{-5} & 3.39\!\times\! 10^{-2} \\
57.0 & 0.160 & 2.89 & 1.55 & 0.374 & 1.73 & 3.39\!\times\! 10^{-2} & 15.7
\end{array}\!\right) 
\end{equation*}

and for the $\pi^-$ fit the covariance matrix is:

\begin{equation*}
\left(\!\begin{array}{cccccccc}
197. & 0.912 & 8.21 & 2.03 & 1.40 & 6.33 & 7.68\!\times\! 10^{-2} & 38.3 \\
0.912 & 4.56\!\times\! 10^{-3} & 3.82\!\times\! 10^{-2} & 9.50\!\times\! 10^{-3} & 6.49\!\times\! 10^{-3} & 2.97\!\times\! 10^{-2} & 3.58\!\times\! 10^{-4} & 0.178 \\
8.21 & 3.82\!\times\! 10^{-2} & 0.345 & 8.52\!\times\! 10^{-2} & 5.85\!\times\! 10^{-2} & 0.265 & 3.22\!\times\! 10^{-3} & 1.60 \\
2.03 & 9.50\!\times\! 10^{-3} & 8.52\!\times\! 10^{-2} & 2.14\!\times\! 10^{-2} & 1.45\!\times\! 10^{-2} & 6.56\!\times\! 10^{-2} & 7.98\!\times\! 10^{-4} & 0.398 \\
1.40 & 6.49\!\times\! 10^{-3} & 5.85\!\times\! 10^{-2} & 1.45\!\times\! 10^{-2} & 1.00\!\times\! 10^{-2} & 4.52\!\times\! 10^{-2} & 5.48\!\times\! 10^{-4} & 0.273 \\
6.33 & 2.97\!\times\! 10^{-2} & 0.265 & 6.56\!\times\! 10^{-2} & 4.52\!\times\! 10^{-2} & 0.208 & 2.49\!\times\! 10^{-3} & 1.24 \\
7.68\!\times\! 10^{-2} & 3.58\!\times\! 10^{-4} & 3.22\!\times\! 10^{-3} & 7.98\!\times\! 10^{-4} & 5.48\!\times\! 10^{-4} & 2.49\!\times\! 10^{-3} & 3.04\!\times\! 10^{-5} & 1.50\!\times\! 10^{-2} \\
38.3 & 0.178 & 1.60 & 0.398 & 0.273 & 1.24 & 1.50\!\times\! 10^{-2} & 7.54
\end{array}\!\right) 
\end{equation*}
\end{widetext}

\begin{figure*}
\mbox{
\begin{minipage}{0.5\textwidth}
\includegraphics[width=9.cm]{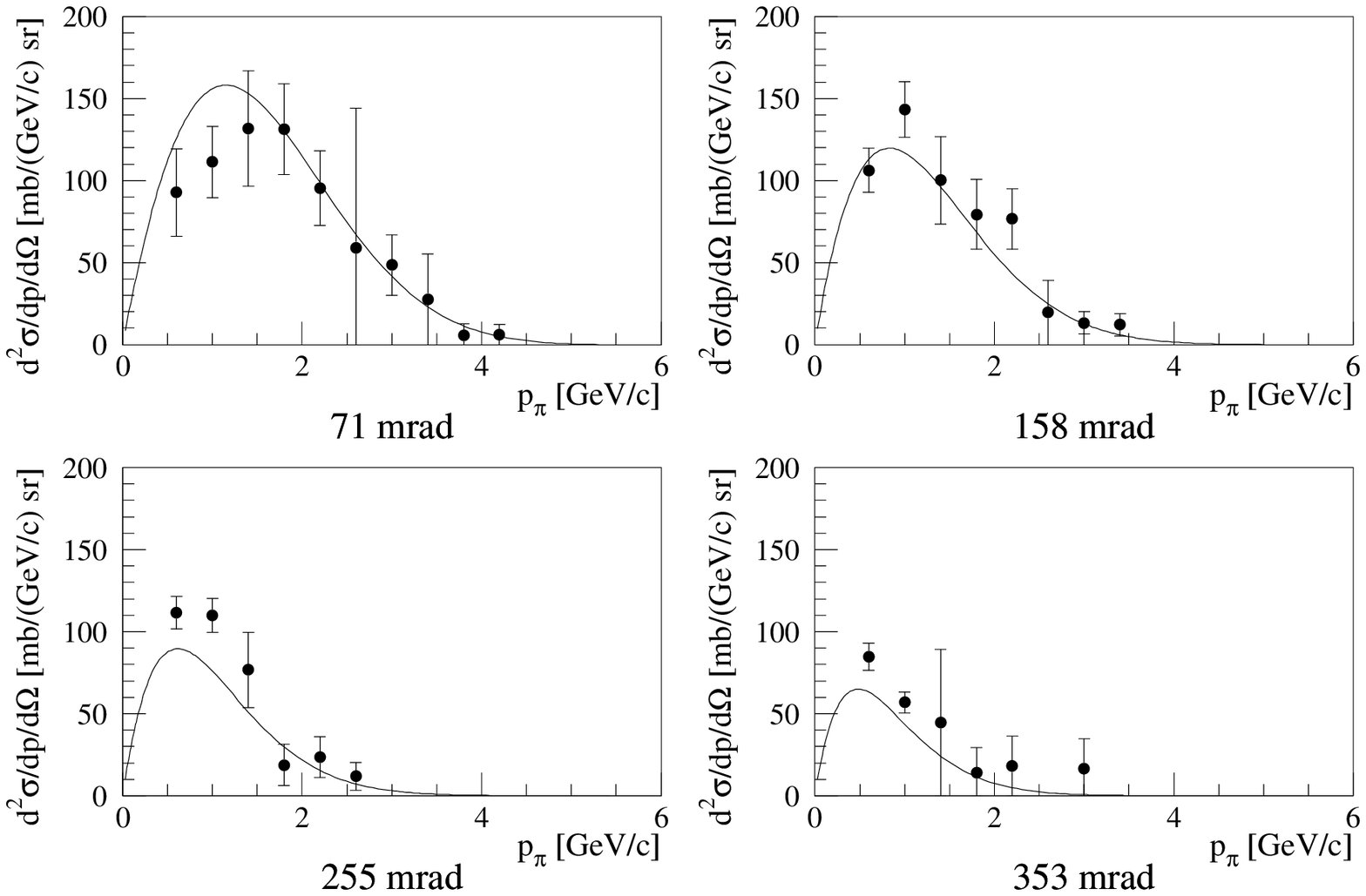}
\caption{\label{plot_6pp} Inclusive p-Be $\pi^+$ production cross section data and fits vs. $\pi^+$ momentum, at 6.4 GeV/c incident proton momentum.  Fits are defined in Table \ref{swfit_1}.}
\end{minipage}
\hspace{0.5cm}
\begin{minipage}{0.5\textwidth}
\includegraphics[width=9.cm]{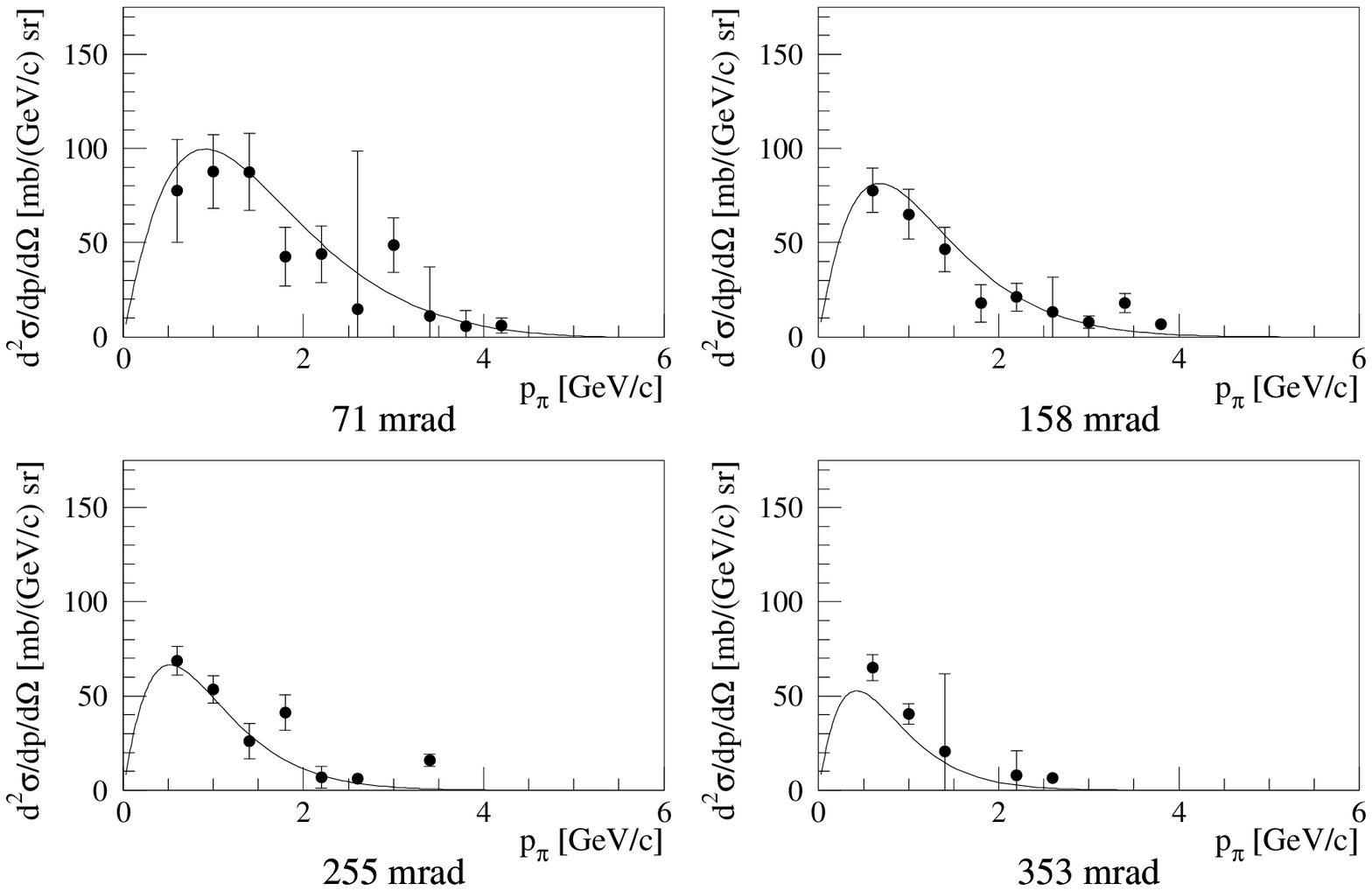}
\caption{\label{plot_6pm}Inclusive p-Be $\pi^-$ production cross section data and fits vs. $\pi^-$ momentum, at 6.4 GeV/c incident proton momentum.  Fits are defined in Table \ref{swfit_1}.}
\end{minipage}
}
\end{figure*}

\begin{figure*}
\mbox{
\begin{minipage}{0.5\textwidth}
\includegraphics[width=9.cm]{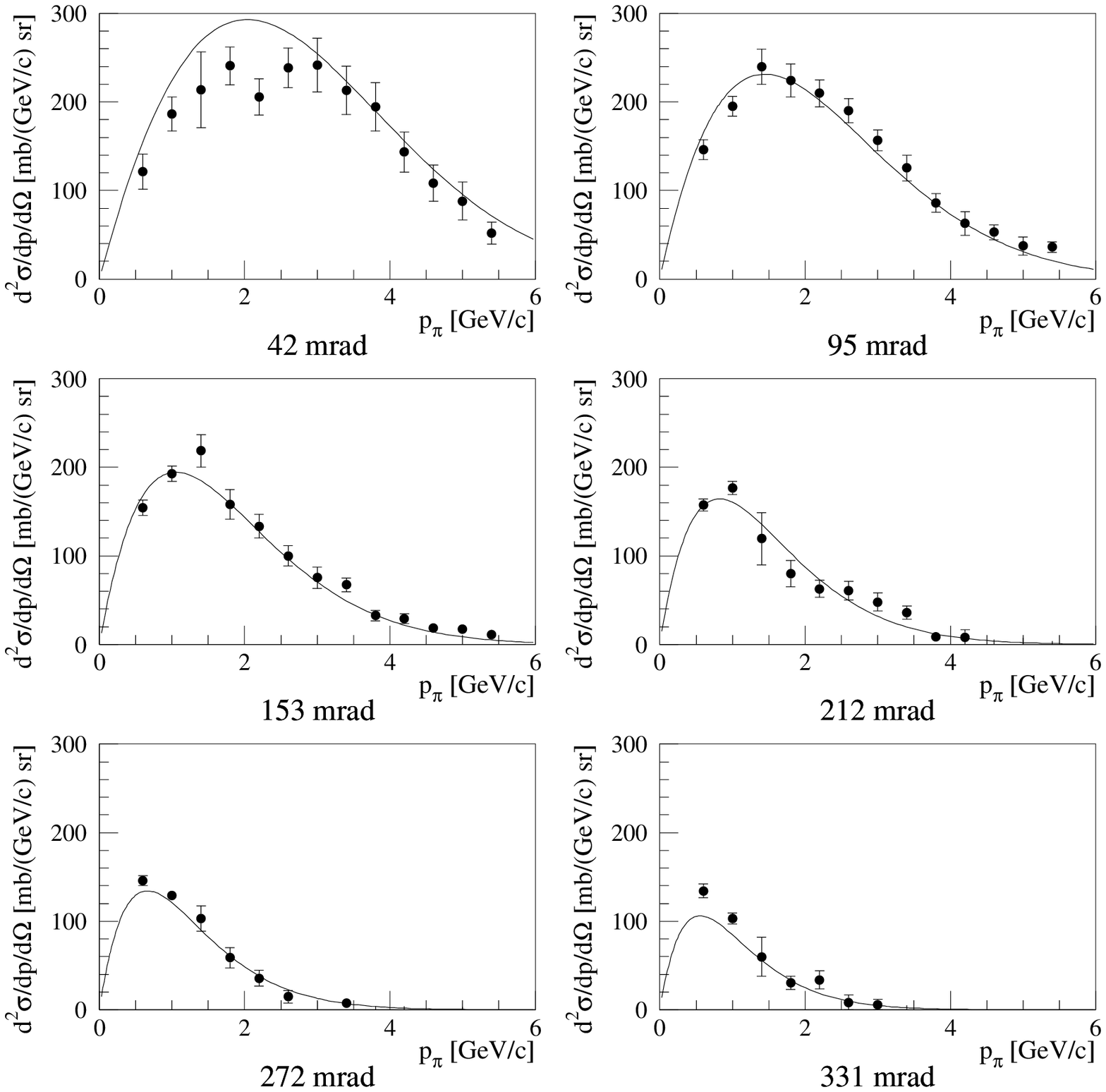}
\caption{\label{plot_12pp} Inclusive p-Be $\pi^+$ production cross section data and fits vs. $\pi^+$ momentum, at 12.3 GeV/c incident proton momentum.  Fits are defined in Table \ref{swfit_1}.}
\end{minipage}
\hspace{0.5cm}
\begin{minipage}{0.5\textwidth}
\includegraphics[width=9.cm]{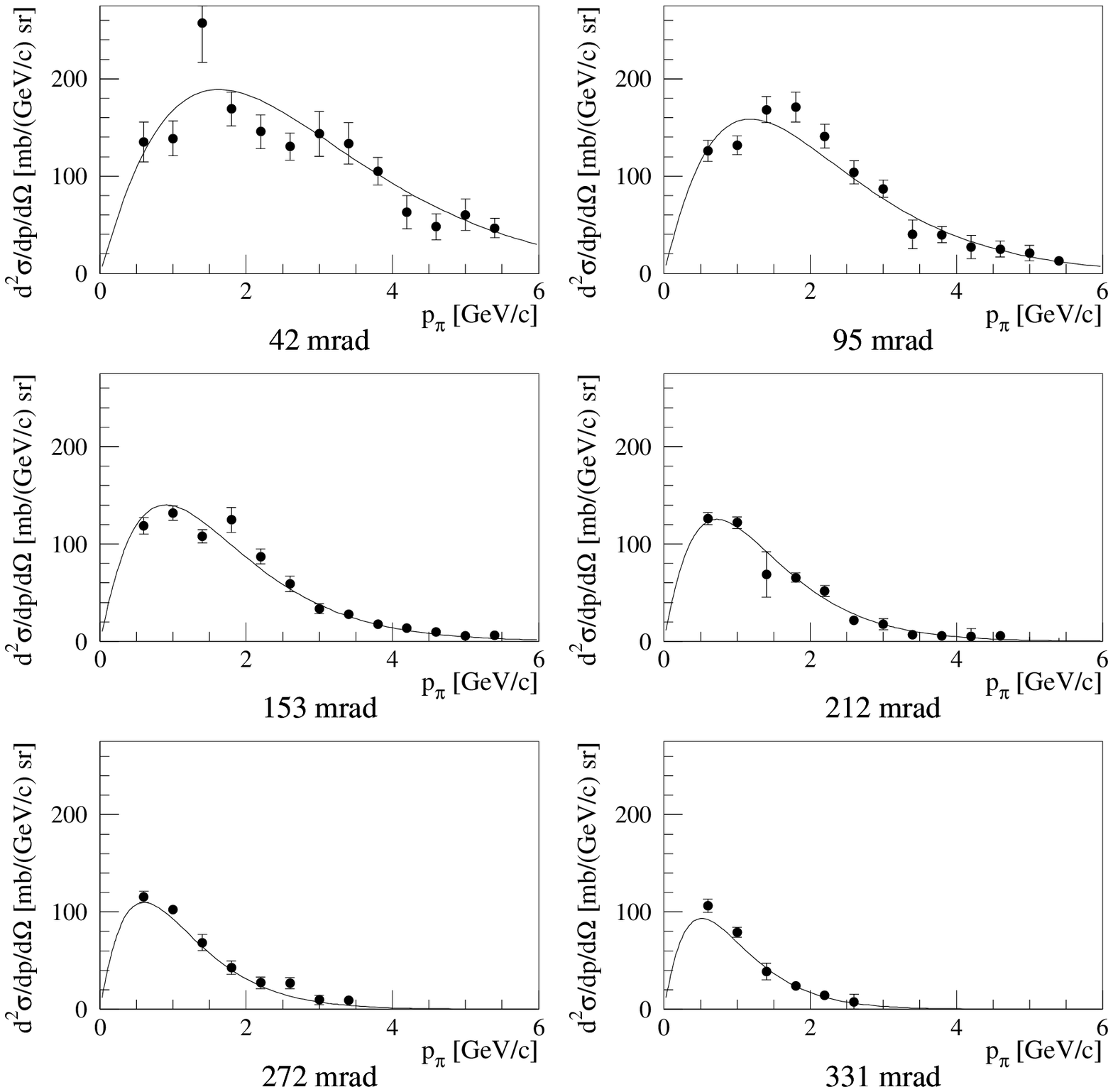}
\caption{\label{plot_12pm}Inclusive p-Be $\pi^-$ production cross section data and fits vs. $\pi^-$ momentum, at 12.3 GeV/c incident proton momentum.  Fits are defined in Table \ref{swfit_1}.}
\end{minipage}
}
\end{figure*}

\begin{figure*}
\mbox{
\begin{minipage}{0.5\textwidth}
\includegraphics[width=9.cm]{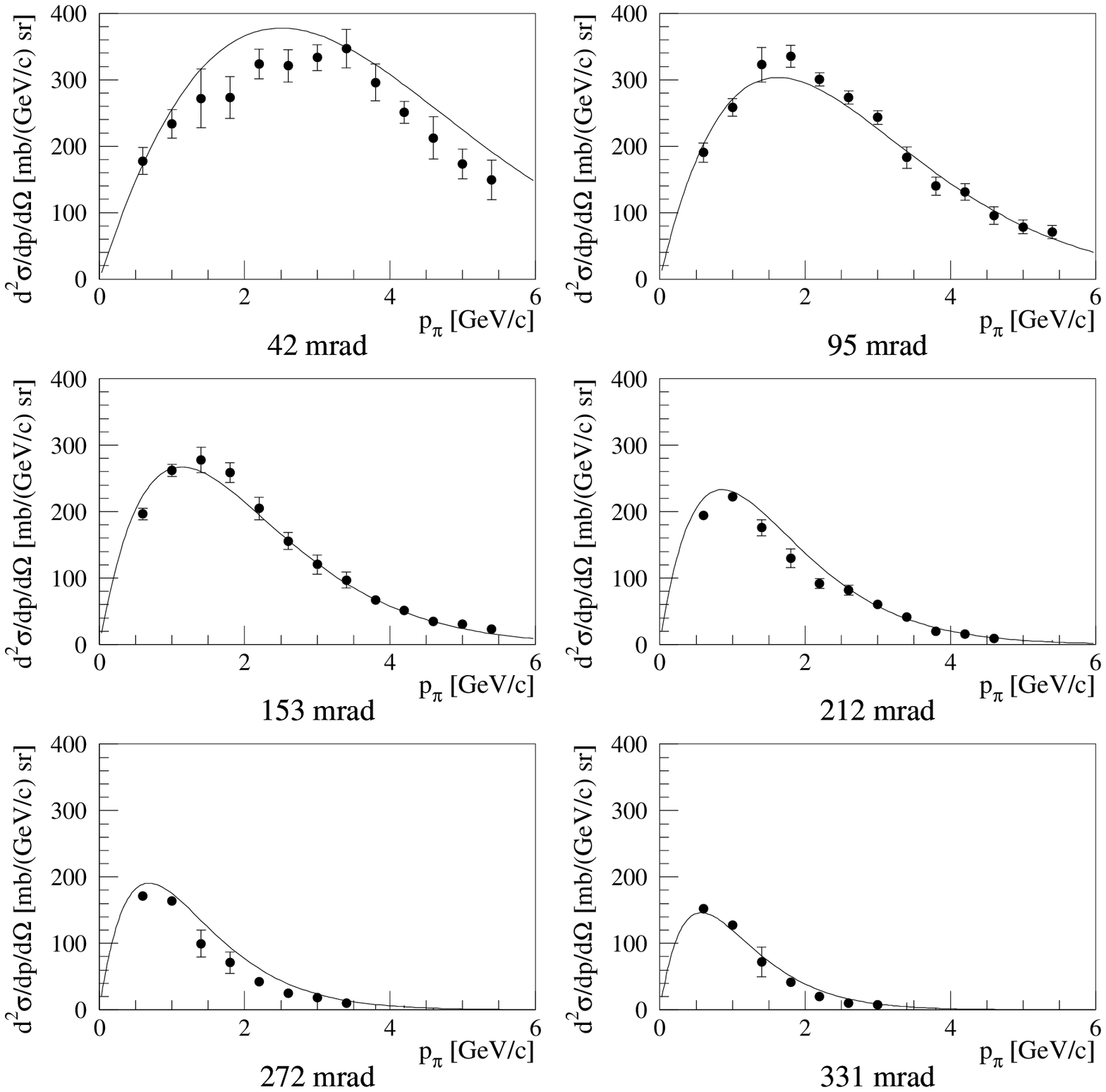}
\caption{\label{plot_17pp} Inclusive p-Be $\pi^+$ production cross section data and fits vs. $\pi^+$ momentum, at 17.5 GeV/c incident proton momentum.  Fits are defined in Table \ref{swfit_1}.}
\end{minipage}
\hspace{0.5cm}
\begin{minipage}{0.5\textwidth}
\includegraphics[width=9.cm]{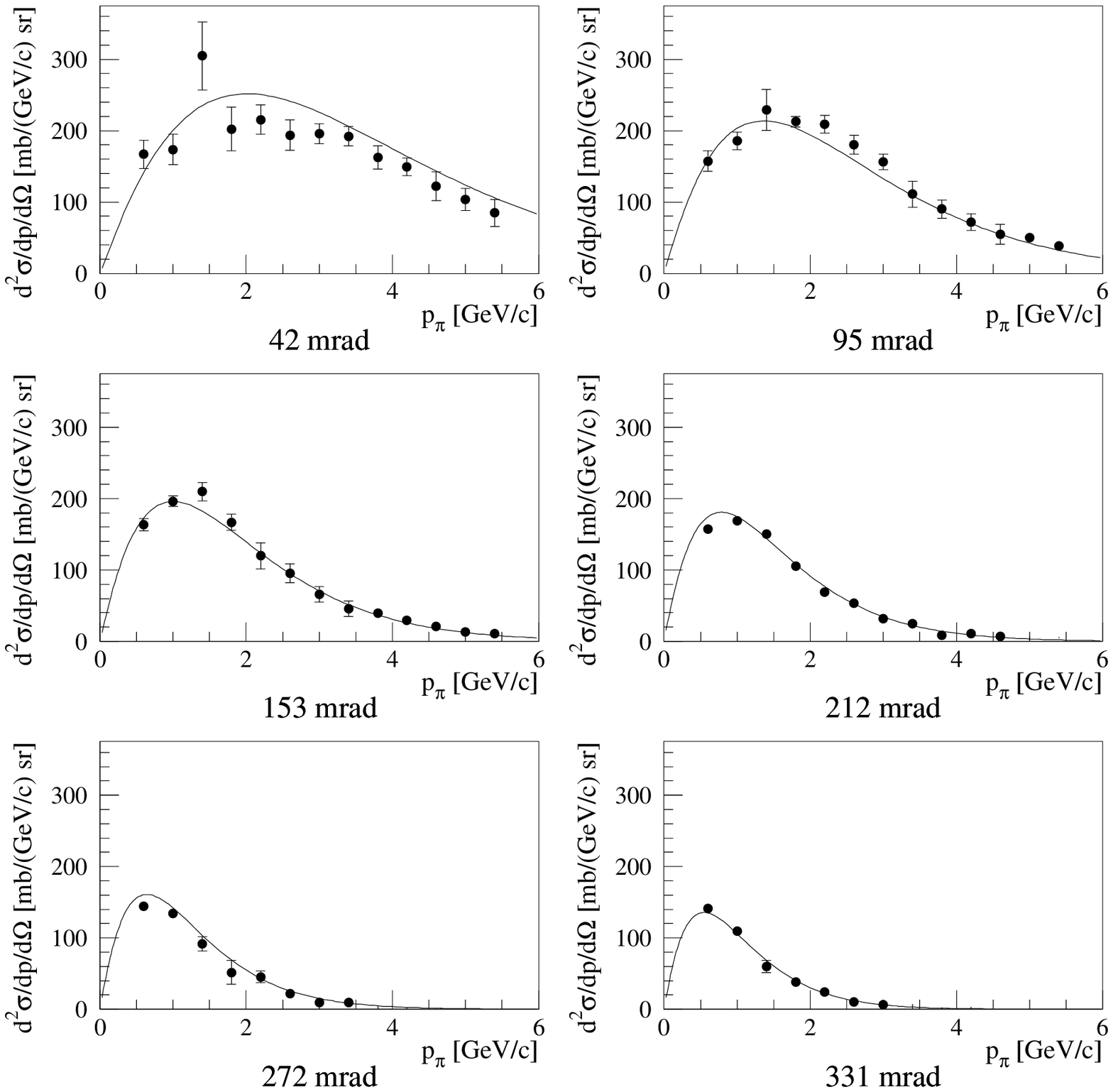}
\caption{\label{plot_17pm}Inclusive p-Be $\pi^-$ production cross section data and fits vs. $\pi^-$ momentum, at 17.5 GeV/c incident proton momentum.  Fits are defined in Table \ref{swfit_1}.}
\end{minipage}
}
\end{figure*}

\noindent Figures \ref{plot_6pp}, \ref{plot_12pp}, and \ref{plot_17pp} show the 6.4, 12.3, and 17.5 GeV/c proton beam momentum $\pi^+$ data sets compared with the fit result overlaid.  The corresponding distributions for $\pi^-$ data are shown in Figures \ref{plot_6pm}, \ref{plot_12pm}, and \ref{plot_17pm}.

Fitting the data to a parameterization like Sanford-Wang allows for the calculation of the pion production cross section as a function of incident proton momentum and for a wide range of pion angles and momenta.  In particular, this fit can be used to generate the primary pion production for experiments such as MiniBooNE with a primary beam momentum (8.9~GeV/c) that was not directly studied, and the fit covariance matrix can be used to calculate the uncertainty in that primary production model. 

The fit is in reasonably good agreement with the data in all but the lowest angular bin (42~mrad).  In this region the fit appears to be systematically above the data, especially in the case of 17.5~GeV/c protons and 12.3~GeV/c protons with secondary  $\pi$ momenta in the 1 to 3~GeV/c range.  This may be due a deficiency in the Sanford-Wang parameterization, or it could be from a systematic and unaccounted for inefficiency in the low angle region.  

\begin{table*}[tb]
\begin{tabular}{@{\extracolsep{0.35cm}}lcrrrrccrrrr} \hline \hline
$\theta$ (mrad) & $p$ (GeV/$c^2$) & \multicolumn{2}{c}{$\pi^+$} & \multicolumn{2}{c}{$\pi^-$} & 
$\theta$ (mrad) & $p$ (GeV/$c^2$) & \multicolumn{2}{c}{$\pi^+$} & \multicolumn{2}{c}{$\pi^-$}\\ \cline{3-4}\cline{5-6}\cline{9-10}\cline{11-12}
 & & $\frac{d^2\sigma}{dpd\Omega}$ & Error & $\frac{d^2\sigma}{dpd\Omega}$ & Error &
 & & $\frac{d^2\sigma}{dpd\Omega}$ & Error & $\frac{d^2\sigma}{dpd\Omega}$ & Error \\ \vspace{-2.5mm}
 71 & 0.6 &  92.7 & 26.8 &  77.5 & 27.2 &  158 & 0.6 & 106.2 & 13.4 &  77.7 & 11.7 \\
    & 1.0 & 111.3 & 21.8 &  87.8 & 19.5 &      & 1.0 & 143.3 & 16.9 &  65.1 & 13.1 \\
    & 1.4 & 131.6 & 35.2 &  87.5 & 20.4 &      & 1.4 & 100.1 & 26.8 &  46.5 & 11.7 \\
    & 1.8 & 131.4 & 27.7 &  42.4 & 15.5 &      & 1.8 &  79.3 & 21.2 &  17.9 & 10.0 \\
    & 2.2 &  95.3 & 22.7 &  43.8 & 14.9 &      & 2.2 &  76.7 & 18.4 &  21.0 &  7.4 \\
    & 2.6 &  59.0 & 85.2 &  14.7 & 84.0 &      & 2.6 &  19.6 & 19.6 &  13.4 & 18.2 \\
    & 3.0 &  48.5 & 18.3 &  48.6 & 14.4 &      & 3.0 &  13.2 &  6.8 &   7.9 &  3.2 \\
    & 3.4 &  27.6 & 27.7 &  11.2 & 25.7 &      & 3.4 &  12.1 &  6.7 &  17.9 &  5.0 \\
    & 3.8 &   5.6 &  7.2 &   5.7 &  8.1 &      & 3.8 &   --- &  --- &   6.6 &  2.7 \\
    & 4.2 &   6.0 &  6.1 &   6.0 &  4.1 &      &     &       &      &       &      \\ \vspace{-2.5mm}
\\
255 & 0.6 & 111.6 &  9.9 &  68.7 &  7.6 &  353 & 0.6 &  84.9 &  8.3 &  65.1 &  6.9 \\
    & 1.0 & 109.9 & 10.3 &  53.5 &  7.2 &      & 1.0 &  56.9 &  6.5 &  40.5 &  5.4 \\
    & 1.4 &  76.8 & 22.9 &  26.0 &  9.3 &      & 1.4 &  44.5 & 44.6 &  20.7 & 41.1 \\
    & 1.8 &  18.7 & 12.8 &  41.3 &  9.4 &      & 1.8 &  14.0 & 15.5 &   --- &  --- \\
    & 2.2 &  23.5 & 12.5 &   7.0 &  5.8 &      & 2.2 &  18.1 & 18.2 &   8.1 & 12.8 \\
    & 2.6 &  11.9 &  8.6 &   6.3 &  2.1 &      & 2.6 &   --- &  --- &   6.6 &  2.9 \\
    & 3.0 &   --- &  --- &   --- &  --- &      & 3.0 &  16.6 & 17.9 &   --- &  --- \\
    & 3.4 &   --- &  --- &  15.9 &  3.2 &      &     &       &      &       &      \\  
\hline
\end{tabular}
\caption{\label{data_6}Pion production cross sections for 6.4 GeV/c protons on Be.}
\end{table*}

\begin{table*}[bt]
\begin{tabular}{@{\extracolsep{0.35cm}}lcrrrrccrrrr} \hline \hline
$\theta$ (mrad) & $p$ (GeV/$c^2$) & \multicolumn{2}{c}{$\pi^+$} & \multicolumn{2}{c}{$\pi^-$} & 
$\theta$ (mrad) & $p$ (GeV/$c^2$) & \multicolumn{2}{c}{$\pi^+$} & \multicolumn{2}{c}{$\pi^-$}\\ \cline{3-4}\cline{5-6}\cline{9-10}\cline{11-12}
 & & $\frac{d^2\sigma}{dpd\Omega}$ & Error & $\frac{d^2\sigma}{dpd\Omega}$ & Error &
 & & $\frac{d^2\sigma}{dpd\Omega}$ & Error & $\frac{d^2\sigma}{dpd\Omega}$ & Error \\ \hline
 42 & 0.6 & 121.1 & 19.9 & 135.1 & 20.4 &   95 & 0.6 & 146.0 & 11.4 & 126.0 & 11.0 \\ 
    & 1.0 & 186.3 & 19.3 & 138.6 & 18.0 &      & 1.0 & 195.0 & 11.3 & 131.5 &  9.8 \\ 
    & 1.4 & 213.6 & 43.0 & 257.1 & 40.2 &      & 1.4 & 239.7 & 19.6 & 168.2 & 13.4 \\ 
    & 1.8 & 240.8 & 21.4 & 168.9 & 17.4 &      & 1.8 & 224.2 & 18.5 & 170.8 & 15.3 \\ 
    & 2.2 & 205.6 & 20.6 & 145.6 & 17.5 &      & 2.2 & 209.9 & 15.2 & 141.0 & 12.3 \\ 
    & 2.6 & 238.5 & 22.4 & 130.2 & 14.1 &      & 2.6 & 190.2 & 13.4 & 103.9 & 11.8 \\ 
    & 3.0 & 241.6 & 30.7 & 143.5 & 23.1 &      & 3.0 & 156.7 & 11.6 &  86.8 &  8.8 \\ 
    & 3.4 & 213.2 & 27.2 & 133.6 & 21.3 &      & 3.4 & 125.5 & 14.7 &  40.1 & 14.8 \\ 
    & 3.8 & 194.6 & 27.2 & 104.7 & 14.2 &      & 3.8 &  86.1 & 10.5 &  39.8 &  8.4 \\ 
    & 4.2 & 143.4 & 22.7 &  62.8 & 17.3 &      & 4.2 &  62.9 & 13.4 &  27.3 & 11.9 \\ 
    & 4.6 & 108.3 & 20.5 &  47.8 & 13.6 &      & 4.6 &  52.8 &  8.3 &  25.0 &  8.1 \\ 
    & 5.0 &  87.9 & 21.4 &  60.2 & 16.2 &      & 5.0 &  37.6 & 10.2 &  20.7 &  8.0 \\ 
    & 5.4 &  51.8 & 12.6 &  46.7 & 10.1 &      & 5.4 &  36.1 &  5.6 &  12.9 &  3.3 \\ \vspace{-2.5mm}
\\    
153 & 0.6 & 154.4 &  8.7 & 118.9 &  8.5 &  212 & 0.6 & 157.3 &  6.8 & 126.3 &  6.3 \\
    & 1.0 & 192.9 &  8.8 & 131.7 &  7.3 &      & 1.0 & 176.5 &  7.4 & 121.9 &  5.8 \\
    & 1.4 & 218.7 & 18.3 & 107.9 &  6.8 &      & 1.4 & 119.6 & 29.5 &  68.6 & 23.2 \\
    & 1.8 & 157.9 & 16.7 & 124.8 & 12.7 &      & 1.8 &  79.8 & 14.8 &  65.6 &  5.1 \\
    & 2.2 & 133.5 & 13.4 &  87.2 &  7.9 &      & 2.2 &  62.6 &  9.6 &  51.6 &  5.6 \\
    & 2.6 & 100.0 & 11.7 &  59.2 &  7.9 &      & 2.6 &  60.6 & 10.5 &  21.3 &  3.6 \\
    & 3.0 &  75.4 & 11.9 &  33.6 &  5.3 &      & 3.0 &  47.9 & 10.2 &  17.5 &  5.5 \\
    & 3.4 &  67.3 &  7.8 &  28.0 &  3.3 &      & 3.4 &  35.8 &  7.5 &   7.0 &  3.6 \\
    & 3.8 &  32.5 &  5.7 &  17.5 &  3.3 &      & 3.8 &   8.5 &  4.6 &   5.5 &  2.6 \\
    & 4.2 &  29.0 &  5.5 &  13.6 &  3.0 &      & 4.2 &   8.2 &  8.2 &   5.2 &  7.7 \\
    & 4.6 &  18.5 &  5.2 &   9.4 &  3.4 &      & 4.6 &   2.8 &  2.9 &   5.4 &  2.7 \\
    & 5.0 &  17.5 &  4.1 &   5.6 &  2.3 &      & 5.0 &   3.6 &  3.7 &   1.3 &  3.4 \\
    & 5.4 &  11.1 &  3.3 &   6.1 &  1.6 &      & 5.4 &   1.9 &  1.9 &   2.1 &  1.4 \\ \vspace{-2.5mm}
\\
272 & 0.6 & 146.0 &  5.4 & 115.5 &  5.5 &  331 & 0.6 & 134.3 &  7.6 & 106.2 &  6.9 \\
    & 1.0 & 129.2 &  4.8 & 102.5 &  4.6 &      & 1.0 & 103.0 &  6.0 &  79.0 &  5.2 \\
    & 1.4 & 103.2 & 14.4 &  68.5 &  8.1 &      & 1.4 &  60.0 & 21.9 &  38.6 &  8.4 \\
    & 1.8 &  58.7 & 11.6 &  42.8 &  7.0 &      & 1.8 &  30.6 &  7.4 &  23.7 &  4.6 \\
    & 2.2 &  35.7 &  8.8 &  27.2 &  6.2 &      & 2.2 &  33.9 & 10.0 &  14.3 &  3.1 \\
    & 2.6 &  14.8 &  7.2 &  27.0 &  5.6 &      & 2.6 &   8.5 &  8.6 &   7.7 &  7.6 \\
    & 3.0 &   4.2 &  4.9 &   9.6 &  4.9 &      & 3.0 &   5.8 &  5.8 &   --- &  --- \\
    & 3.4 &   7.3 &  3.6 &   9.2 &  1.8 &      & 3.4 &   2.5 &  2.7 &   --- &  --- \\
    & 3.8 &   1.7 &  2.3 &   --- &  --- &      &     &       &      &       &      \\
    & 4.2 &   1.7 &  2.2 &   --- &  --- &      &     &       &      &       &      \\
    & 5.0 &   1.5 &  2.0 &   --- &  --- &      &     &       &      &       &      \\
\hline
\end{tabular}
\caption{\label{data_12}Pion production cross sections for 12.3 GeV/c protons on Be.}
\end{table*}

\begin{table*}[bt]
\begin{tabular}{@{\extracolsep{0.35cm}}lcrrrrccrrrr} \hline \hline
$\theta$ (mrad) & $p$ (GeV/$c^2$) & \multicolumn{2}{c}{$\pi^+$} & \multicolumn{2}{c}{$\pi^-$} & 
$\theta$ (mrad) & $p$ (GeV/$c^2$) & \multicolumn{2}{c}{$\pi^+$} & \multicolumn{2}{c}{$\pi^-$}\\ \cline{3-4}\cline{5-6}\cline{9-10}\cline{11-12}
 & & $\frac{d^2\sigma}{dpd\Omega}$ & Error & $\frac{d^2\sigma}{dpd\Omega}$ & Error &
 & & $\frac{d^2\sigma}{dpd\Omega}$ & Error & $\frac{d^2\sigma}{dpd\Omega}$ & Error \\ \hline
 42 & 0.6 & 177.7 & 20.0 & 166.9 & 19.8 &  95 & 0.6 & 190.5 & 14.3 & 157.3 & 14.2 \\
    & 1.0 & 233.4 & 21.6 & 173.5 & 21.2 &     & 1.0 & 258.5 & 13.1 & 185.7 & 12.8 \\
    & 1.4 & 271.8 & 44.3 & 305.0 & 47.8 &     & 1.4 & 322.6 & 26.2 & 229.0 & 28.5 \\
    & 1.8 & 273.4 & 31.6 & 202.3 & 30.8 &     & 1.8 & 335.2 & 16.3 & 212.7 &  7.4 \\
    & 2.2 & 323.4 & 22.3 & 215.4 & 20.6 &     & 2.2 & 300.9 &  9.9 & 209.2 & 12.2 \\
    & 2.6 & 321.1 & 24.3 & 193.9 & 21.5 &     & 2.6 & 273.6 &  9.8 & 180.3 & 13.3 \\
    & 3.0 & 333.5 & 19.6 & 196.0 & 14.1 &     & 3.0 & 243.2 & 10.1 & 156.2 & 10.5 \\
    & 3.4 & 347.0 & 29.1 & 192.0 & 13.6 &     & 3.4 & 183.0 & 16.2 & 111.1 & 18.2 \\
    & 3.8 & 295.9 & 27.7 & 162.4 & 16.0 &     & 3.8 & 139.9 & 13.6 &  90.1 & 12.5 \\
    & 4.2 & 250.9 & 16.7 & 149.4 & 12.3 &     & 4.2 & 131.4 & 12.4 &  71.4 & 11.6 \\
    & 4.6 & 212.4 & 31.6 & 122.0 & 20.0 &     & 4.6 &  95.7 & 13.0 &  54.6 & 14.1 \\
    & 5.0 & 173.5 & 22.2 & 103.9 & 15.6 &     & 5.0 &  78.3 & 10.4 &  50.0 &  6.3 \\
    & 5.4 & 149.4 & 29.6 &  84.6 & 18.7 &     & 5.4 &  70.7 &  9.6 &  38.5 &  6.6 \\ \vspace{-2.5mm} \\
153 & 0.6 & 196.3 &  8.6 & 163.8 &  8.5 & 212 & 0.6 & 194.0 &  6.3 & 157.6 &  5.9 \\
    & 1.0 & 261.9 &  9.2 & 196.2 &  7.3 &     & 1.0 & 222.0 &  6.2 & 168.8 &  5.7 \\
    & 1.4 & 277.7 & 18.9 & 209.7 & 12.7 &     & 1.4 & 175.6 & 12.3 & 150.4 &  6.0 \\
    & 1.8 & 258.6 & 15.0 & 166.9 & 11.5 &     & 1.8 & 129.9 & 13.9 & 105.5 &  6.6 \\
    & 2.2 & 204.9 & 17.0 & 119.8 & 18.0 &     & 2.2 &  91.7 &  7.4 &  69.2 &  4.1 \\
    & 2.6 & 155.7 & 12.5 &  95.2 & 13.0 &     & 2.6 &  81.7 &  7.4 &  53.5 &  3.3 \\
    & 3.0 & 120.4 & 14.3 &  65.9 & 11.1 &     & 3.0 &  60.2 &  6.2 &  32.0 &  2.7 \\
    & 3.4 &  97.1 & 12.2 &  45.9 & 11.0 &     & 3.4 &  40.9 &  6.3 &  24.6 &  3.2 \\
    & 3.8 &  67.2 &  6.5 &  39.3 &  5.0 &     & 3.8 &  20.1 &  7.0 &   8.4 &  5.9 \\
    & 4.2 &  51.5 &  5.0 &  29.3 &  3.9 &     & 4.2 &  15.7 &  2.7 &  10.5 &  0.9 \\
    & 4.6 &  34.9 &  6.1 &  21.0 &  3.7 &     & 4.6 &	9.4 &  2.0 &   6.9 &  2.4 \\
    & 5.0 &  30.1 &  3.7 &  13.1 &  3.2 &     & 5.0 &	5.4 &  1.6 &   4.2 &  2.3 \\
    & 5.4 &  23.2 &  3.7 &  10.8 &  1.4 &     & 5.4 &	4.8 &  1.7 &   5.4 &  0.9 \\  \vspace{-2.5mm} \\
272 & 0.6 & 171.0 &  4.7 & 144.2 &  5.9 & 331 & 0.6 & 152.4 &  6.6 & 141.4 &  5.9 \\
    & 1.0 & 163.6 &  3.7 & 134.5 &  3.2 &     & 1.0 & 127.1 &  5.5 & 109.3 &  5.2 \\
    & 1.4 &  99.6 & 20.0 &  91.4 & 10.0 &     & 1.4 &  72.4 & 22.4 &  59.6 &  8.6 \\
    & 1.8 &  71.2 & 16.0 &  51.5 & 16.6 &     & 1.8 &  41.2 &  4.9 &  38.3 &  3.0 \\
    & 2.2 &  42.2 &  6.8 &  45.4 &  7.9 &     & 2.2 &  20.1 &  3.8 &  23.9 &  2.4 \\
    & 2.6 &  24.8 &  5.0 &  21.5 &  6.2 &     & 2.6 &  10.1 &  3.3 &   9.8 &  1.9 \\
    & 3.0 &  18.1 &  3.9 &   9.6 &  2.7 &     & 3.0 &	7.4 &  4.0 &   6.7 &  1.5 \\
    & 3.4 &   9.8 &  2.7 &   9.8 &  2.0 &     & 3.4 &	5.6 &  5.6 &   2.0 &  5.0 \\
    & 3.8 &   4.3 &  3.0 &   1.6 &  3.0 &     & 3.8 &	0.9 &  1.1 &   1.1 &  1.4 \\
    & 4.2 &   4.7 &  2.2 &   0.7 &  1.2 &     & 4.2 &	--- &  --- &   1.3 &  2.1 \\
    & 4.6 &   0.7 &  1.1 &   0.8 &  0.4 &     &     &	    &	   &	   &	  \\
    & 5.0 &   0.9 &  0.9 &   0.6 &  0.6 &     &     &	    &	   &	   &	  \\
    & 5.4 &   0.1 &  0.3 &   --- &  --- &     &     &	    &	   &	   &	  \\  
\hline \hline
\end{tabular}
\caption{\label{data_17}Pion production cross sections for 17.5 GeV/c protons on Be.}
\end{table*}

\begin{acknowledgments}
We wish to thank R.~Hackenburg and the MPS staff, J.~Scaduto and G.~Bunce for their support during E910 data taking.  This work has been supported by the U.S. Department of Energy under contracts with BNL (DE-AC02-98CH10886), Columbia (DE-FG02-86ER40281), ISU (DOE-FG02-92ER4069), KSU (DE-FG02-89ER40531), LBNL (DE-AC03-76F00098), LLNL (W-7405-ENG-48), ORNL (DE-AC05-96OR22464), and UT (DE-FG02-96ER40982), and the National Science Foundation under contract with Columbia (PHY-0098826), and Florida State University (PHY-9523974).
\end{acknowledgments}

\bibliography{e910}

\end{document}